\documentstyle[editedvolume]{crckapb} 
\include{psfig}

\begin{opening}
\title{Globular Clusters in Blue Compact Galaxies 
as Tracers of the Starburst 
History}

\author{G. \"Ostlin}
\institute{Institut d'Astrophysique de Paris\\
           98 bis Boulevard Arago \\
	   75014 Paris, France}

\end{opening}

\runningtitle{Globular Clusters in Blue Compact Galaxies}

\begin{document}


\begin{abstract}

Representing single stellar populations, globular clusters (GCs) are relatively easy to model, thus providing powerful tools for studying  the evolution of galaxies. This has been demonstrated for the blue compact galaxy ESO338-IG04. GC systems in galaxies may be fossils of starbursts and  mergers. Thus studies of GCs in the local universe may add to our understanding of the formation and evolution of galaxies and the distant universe.

\end{abstract}

\section{Introduction -- Globular cluster formation and destruction}

Globular clusters (GCs) are in general old stellar systems with masses $10^4$ to $10^7 M\odot$, and are believed to be among the first ingredients to form in the process of galaxy formation. Thus understanding how GCs form is vital for understanding how galaxies form.
Extragalactic GC systems in e.g. elliptical (E) galaxies often have
bimodal colour distributions, indicating the presence of  populations with 
different metallicity and/or age.
Studies of merging galaxies, e.g. the "Antennae" (Whitmore and Schweizer 1995) and NGC 7252 (Miller et al. 1997), have shown that these often contain young 
GC candidates in great numbers. These galaxies are believed to  evolve into E galaxies as the merger remnants relax. The GC candidates in the more evolved  mergers have redder colours indicating higher ages. 
Very young  so-called ``super star clusters'' (SSCs) have been 
found in many starburst galaxies, including dwarfs (e.g. Meurer et al. 1995). 
The SSCs have properties which largely agree with 
those  expected for young GCs, although the masses are quite uncertain, 
and are probably younger examples of the 
objects seen in the mergers. The triggering mechanism for the starbursts in 
the dwarfs that host SSCs is still an open question, but dwarf mergers are 
not ruled out.  SSCs have also been found in the centres and  
circum-nuclear rings in giant barred spirals (e.g. Barth et al. 1995
and Kristen et al. 1997) 

In conclusion  young globular cluster candidates are found in extreme and
energetic environments indicating that they can only be formed under special
conditions. The coeval formation of many GCs will require extreme conditions,
e.g. very high pressures and gas densities (Elmegreen and Efremov, 1997),
conditions which are fulfilled in mergers. Moreover, bars trigger gas flows
and  nuclear rings may be created by dynamical resonances, enhancing the 
density. A newly formed GC will not automatically become an old GC (like the 
ones in our Galaxy) as time goes by, but  faces the risk of destruction 
and dissolution. Their ability to survive depends on their interaction
with their environment, but also on their  IMF. 
A galactic bar for instance is not a favourable place for GC survival since
strong shocks will easily disrupt many young GCs. The conditions in mergers,
and in particular dwarf galaxies may be more favourable for GC survival.

\section{Young and old GCs in ESO~338-IG04}

ESO338-IG04 is a blue compact galaxy (BCG). A BCG is characterised by 
compact appearance and HII region like spectra indicating high star 
formation rates, and in general low chemical abundances. Most BCGs are 
dwarf galaxies that, for some reason, presently are undergoing starbursts. 
HST/WFPC2 images reveals that ESO338-IG04 hosts 
a very rich population of compact star clusters counting more than 
one hundred objects (\"Ostlin et al. 1998).
The centre is crowded with
young SSCs, but in addition lots of intermediate age and old objects 
are found outside the starburst region (there might well be some in the centre  as well 
but they drown in the light from young SSCs).  
Photometric modelling (using Salpeter and Miller-Scalo IMFs) indicates
masses  in the range $10^4$  to more than  $ 10^7
M\odot$ and a wide range of ages, 
from a few Myr up to 10 Gyr (see Fig 1). The spread in ages is real and not caused by observational errors. Moreover, there are 
groupings in the age distribution indicating that the 
cluster formation history has not been continuous. Most apparent is of
course the numerous very young and luminous SSCs, but perhaps 
even more interesting is the presence of one or two populations of 
massive GCs with age 2 to 5 Gyr. This population, when corrected for 
objects below the detection limit, shows that starburst progenitor had 
a specific frequency $S_N$ (a measure of the number of GCs relative to
the host galaxy luminosity) comparable to those of giant Es. In addition 
there are old ($\ge 10$ Gyr) 
and 0.5 Gyr GCs present. 
Follow up spectroscopy of two GC candidates in ESO338-IG04  confirms the 
the results from  photometry (\"Ostlin et al.  1998c, in preparation). The dynamics and morphology
of this galaxy indicates that a dwarf merger is responsible for the
starburst (\"Ostlin et al. 1998a, 1998b). Thus we have a splendid low 
luminosity 
($M_B = -19$) counterpart
of the giant mergers  mentioned above. Even if young GCs of course may 
disrupt or dissolve, the presence of intermediate age GCs proves that 
GC formation is not a phenomenon that was isolated to the very earliest 
days of galaxy formation. SSCs are often found in BCGs which suggest that they might be a good 
laboratory for studying GC formation; and a closer look for aged GCs 
may provide important information on previous bursts.

\begin{figure}
\psfig{file=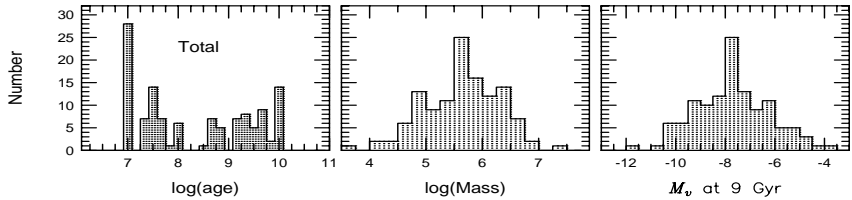,height=3.61cm,width=12.5cm}
\caption[]{Modelled ages (left), masses (middle) and absolute V magnitudes 
when  transforming all objects to an age of 9 Gyr (right). The
results  are based on a standard Salpeter IMF, but are very similar for a 
Miller-Scalo IMF  (\"Ostlin et al. 1998a).}
\end{figure}

\section{GCs as starburst fossils}

We have seen that objects which are likely to be  young GCs have been found in 
galaxies which are starbursts and/or mergers. The presence of intermediate 
age GCs in ESO338-IG04 and  somewhat aged objects in the merger 
remnants suggest that at least a considerable fraction of young GCs will 
survive. It is a general property of starbursts to reveal the 
presence of SSCs when studied at high spatial resolution. R136, the central 
cluster in 30 Dor., may {\it perhaps} be the closest example of a  GC in the making.
Even if not all SSCs become GCs, we can conclude that all newly formed GCs  must have properties similar to SSCs. Moreover the coeval formation of many massive GCs would produce a starburst in itself. Let us illustrate this in a figure (Fig 2): A starburst leads to SSCs of which at least a fraction becomes GCs, 
and the reverse is also true, although SSCs may form also in bars, which 
however are hostile environments. Therefore the age distribution of the GC  
population  can be used to infer the SSC, and thus 
starburst, history. It is also clear that mergers are capable of trigger 
starbursts, but we do not know yet if a merger is required for the occurrence 
of a global starburst in a galaxy. If that would be the case, GC populations 
could be used to trace the merger history of galaxies. Of course, the question
marks in Fig. 2 must be straightened out before GCs can be
used as general probes. 
 
\begin{figure}
\psfig{file=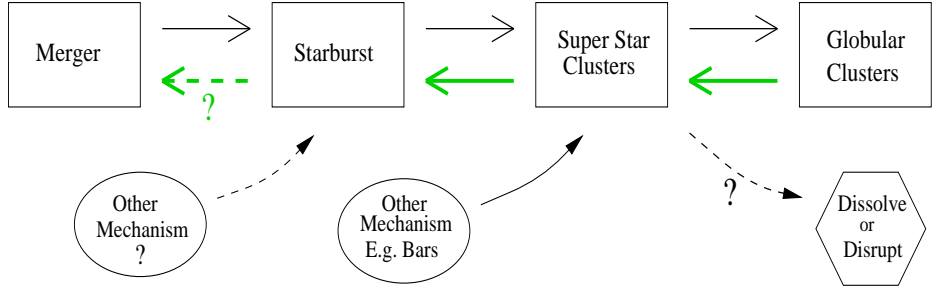,height=3.8cm,width=12.3cm,angle=-90}
\caption[]{A sketch of the connection between GCs, SSCs, starbursts and mergers }
\end{figure}

Although GCs  provide information on the starbursts history of galaxies, they do not 
necessarily tell us about the overall star formation history, 
afterall GC free galaxies exist. There 
is obviously two different modes of star formation: the ``violent'' starburst mode 
which favours cluster formation and the ``quiet'' mode that may still produce 
the bulk of the field stars in most galaxies (Van den Bergh 1998).
In nearby galaxies the star formation history (SFH) of the field population can be studied through deep colour magnitude diagrams (cf. Tolstoy this 
volume). In most galaxies however one has to infer the overall SFH from 
the integrated stellar population. A GC is much easier to model  because 
it represents a true single coeval (on the scale of a few Myrs) stellar 
population. Spectroscopy can be used to investigate metallicities to 
circumvent the age-metallicity degeneracy. Thus GCs can serve as excellent 
probes for unveiling the starburst history in moderately distant galaxies.
But also a more complete picture of the GC content in local galaxies 
would provide important information. 
This might be of importance for interpreting deep surveys which often 
are biased towards starburst galaxies. Depending on how strong the 
connection to mergers will prove to be, GC populations 
may also be useful for studying the morphological and number evolution 
of the galaxy population.

\end{document}